\documentclass{article}
\usepackage[T1]{fontenc}
\usepackage[utf8]{inputenc}
\usepackage{ismir,amsmath,cite,url}
\usepackage{graphicx}
\usepackage{color}
\usepackage{subcaption}
\usepackage{booktabs}
\usepackage{enumitem}

\title{Exploring GPT's Ability as a Judge in Music Understanding}

\multauthor
{Kun Fang$^{1,2}$ \hspace{1cm} Ziyu Wang$^{3,2}$ \hspace{1cm} Gus Xia$^{2,3}$ \hspace{1cm} Ichiro Fujinaga$^{1}$}
{\vspace{-0.3cm} \\
 $^1$ Schulich School of Music, McGill University \hspace{0.5cm}
$^2$ Machine Learning Department, MBZUAI \\
$^3$ Computer Science Department, NYU Shanghai\\
{\tt{kun.fang@mail.mcgill.ca, ziyu.wang@nyu.edu,}} \\
{\tt{gus.xia@mbzuai.ac.ae, ichiro.fujinaga@mcgill.ca}}
}

\def\authorname{K. Fang, Z. Wang, G. Xia, and I. Fujinaga}

\usepackage[bookmarks=false,pdfauthor={\authorname},pdfsubject={\papersubject},hidelinks]{hyperref}

\begin{document}
\maketitle
\begin{abstract}
Recent progress in text-based Large Language Models (LLMs) and their extended ability to process multi-modal sensory data have led us to explore their applicability in addressing music information retrieval (MIR) challenges. In this paper, we use a systematic \textit{prompt engineering} approach for LLMs to solve MIR problems. We convert the music data to symbolic inputs and evaluate LLMs' ability in detecting annotation errors in three key MIR tasks: beat tracking, chord extraction, and key estimation. A \textit{concept augmentation} method is proposed to evaluate LLMs' music reasoning consistency with the provided music concepts in the prompts. Our experiments tested the MIR capabilities of Generative Pre-trained Transformers (GPT). Results show that GPT has an error detection accuracy of 65.20\%, 64.80\%, and 59.72\% in beat tracking, chord extraction, and key estimation tasks, respectively, all exceeding the random baseline. Moreover, we observe a positive correlation between GPT's error finding accuracy and the amount of concept information provided. The current findings based on symbolic music input provide a solid ground for future LLM-based MIR research.\footnote{Code repository: \url{https://github.com/kunfang98927/gpt-eval-mir}}
\end{abstract}

\section{Introduction}\label{sec:introduction}

Recent advancements in text-based Large Language Models (LLMs) have showcased their significant reasoning and knowledge retrieval capabilities across various domains, including music understanding. For instance, the standard GPT-4 model performs better than random on music theory questions \cite{chatmusician}. This success raises the question of whether such text-based reasoning abilities could enhance Music Information Retrieval (MIR) tasks. From a psychological perspective, we are interested in how a \textit{cognition module}, typically represented by a text-based LLM, can possibly interplay with a \textit{perception module}, typically represented by an MIR network, to improve music understanding.

A key challenge in achieving this goal is the inherent difference between music and text modality, which typically requires aligning data in other modalities to text. Common strategies include either transforming all inputs into a unified modality \cite{chatmusician,spae}, or developing adapters tailored to other domains, such as MiniGPT-5 \cite{minigpt} and NextGPT \cite{nextgpt}. Given the substantial data requirements and training costs involved in addressing cross-modality issues, we believe a practical initial step for LLM-based MIR research is to translate sensory inputs into symbolic representations and investigate the performance of text-based LLMs in a training-free way (e.g., prompt engineering \cite{wei2022chain}). This methodology allows us to assess how much cognition alone, without additional auditory perception, can enhance MIR tasks.

To this end, we propose a systematic prompt engineering method to assess the music understanding capabilities of text-based LLMs, focusing specifically on their ability to \textit{detect errors} in MIR annotations. Each task input includes: 1) a music segment converted into MIDI or higher-level musical features, 2) a corresponding MIR annotation with deliberately inserted errors, and 3) a text prompt that introduces the MIR problem and outlines relevant musical concepts. The LLM's role is to pinpoint errors within the musical annotations, effectively acting as an MIR judge. In all the tasks, annotation errors are randomly applied at controlled rates, and prompts are crafted using common prompt engineering techniques. Additionally, we propose a \textit{concept augmentation} strategy to evaluate the LLM's behavioral consistency in response to the musical concepts provided. This involves adjusting the occurrence of certain musical concepts in the prompt, such as replacing a musical term (e.g., pitch sequence) with a more general term (e.g., time series) to obscure a concept, or vice versa, to explore whether these changes influence the LLM's performance in predictable ways.

We carried out experiments using the GPT-3.5 model (hereafter, GPT), targeting three MIR tasks: beat tracking, chord extraction, and key estimation. The experiment results indicate that the error detection rates are higher than random, achieving scores of 65.20\%, 64.80\%, and 59.72\%, respectively. Furthermore, the concept augmentation experiments show that GPT's performance broadly correlates with the amount of musical concepts introduced in the prompts. These findings suggest that GPT exhibits measurable music understanding capability, which sets a foundational baseline for future LLM-based MIR research.
In sum, the contributions of the paper are as follows:

\begin{enumerate}[leftmargin=20pt]\setlength{\itemsep}{0pt}
\vspace{-0.12cm}
\item \textbf{We pioneer the integration of MIR problems with text-based LLMs.} Our approach utilizes prompt-engineering techniques for MIR error detection and adopts the symbolic music format to unify music and text modality, which does not require additional training.
\vspace{-0.12cm}
\item \textbf{We perform a systematic study on GPT’s abilities as a judge} in beat tracking, chord extraction, and key estimation tasks, demonstrating GPT's capability in solving MIR problems.
\vspace{-0.12cm}
\item \textbf{We provide a solid ground for LLM-based MIR research.} The proposed methodology sets a baseline for future studies.

\end{enumerate}

\section{Related Work}

Recently, the advancements of text-based LLMs \cite{gpt4,vicuna,llama} have expanded beyond textual data, incorporating capabilities to interpret information from various other modalities. In the computer music domain, the research to combine text and audio LLMs is also popular. For example, ChatMusician is a text-based LLM, which focuses mainly on generating symbolic music in ABC notation \cite{chatmusician}; MusicGen \cite{musicgen} and Coco-Mulla \cite{lin2023content} are audio-based LLMs allowing text and symbolic music control; and MU-LLaMA is an audio-to-text model for caption generation \cite{Mullama}. Despite all these achievements, the current cross-modal research of text-based LLMs is restricted to generative tasks; and their ability to reason about cross-domain data is still under-researched. The focus of this paper is to evaluate whether LLMs can be used for music understanding and solving MIR problems.

In most cross-modal LLM studies, extensive training is required to align cross-modal information. These approaches involve training separate adapters to align the pre-trained model with other-domain data \cite{nextgpt,minigpt,video}, fine-tuning an LLM on symbolic cross-domain data \cite{chatmusician}, or learning a trainable autoencoder to convert other-domain data to text tokens \cite{spae}. In the music domain, since music can be naturally represented as readable symbolic representations, we propose using prompt engineering methods to connect the music and text domains to avoid extra training.

The cross-domain prompt engineering methods used in this paper originate from the text domain. These strategies involve chain-of-thought \cite{wei2022chain}, few-shot prompting \cite{brown2020language}, least-to-most prompting \cite{zhou2022least}, and many others \cite{wang2022self,zhou2022large,yao2022react}. These methods show that the more organized the prompt is, the better the LLM will be able to reason. To the best of our knowledge, we present the first attempt of using prompt engineering to teach LLMs to reason about music. We aim to explore to what extent music reasoning alone can help MIR.

\begin{figure*}[t!]
    \centering
    \includegraphics[width=0.95\linewidth]{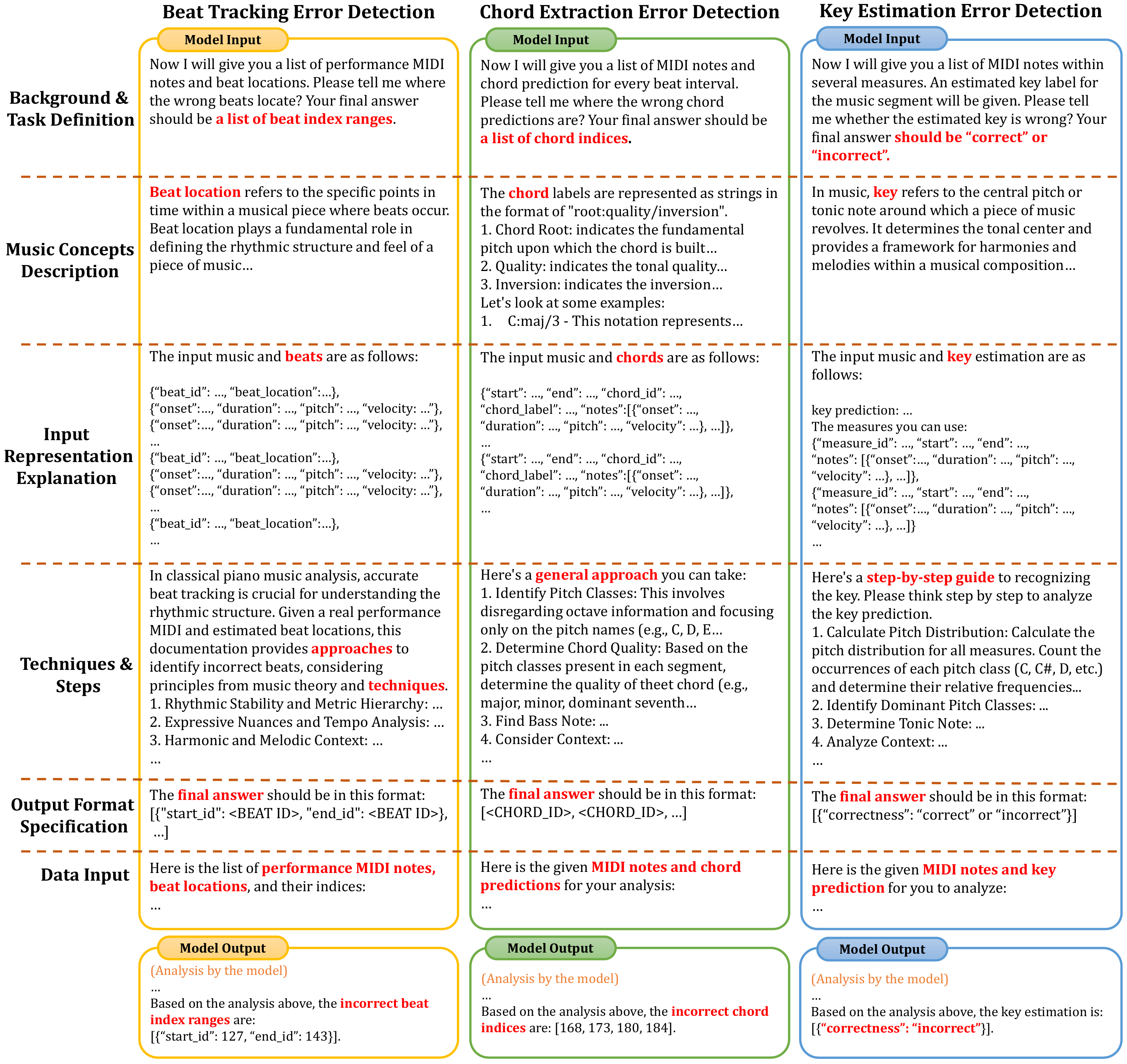}
    \caption{The example prompts and model outputs for the three error d
    etection MIR tasks: beat tracking, chord extraction, and key estimation. Some keywords are highlighted in red in this figure for better readability. Orange texts indicate omitted content. The prompt structure is shown on the left.}
    \label{fig:prompt-structure}
\end{figure*}

\section{Methodology}

In this study, we use prompt engineering to evaluate the capabilities of text-based LLMs through three MIR error detection tasks: beat tracking, chord extraction, and key estimation (as shown in Figure~\ref{fig:prompt-structure}).
In Section~\ref{subsec:3.1}, we introduce the task definition and data representations for each task. In Section~\ref{subsec:3.2}, we discuss the structure and main components of the prompts.
Finally, Section~\ref{subsec:3.3} introduces the proposed concept augmentation methods to test the LLMs' music reasoning ability with respect to the music concepts included in prompts. 

\subsection{Task Definition and Data Representation} \label{subsec:3.1}

For symbolic MIR tasks, beat tracking determines the precise timing of beats in a MIDI-like music representation, chord extraction assigns a chord label to each segment, and key estimation identifies the musical key of each segment. Building on these tasks, we introduce a novel task: MIR error detection. This task involves identifying errors specific to each of the three traditional MIR tasks. The following subsections define the error detection tasks for beat tracking, chord extraction, and key estimation.

\subsubsection{Beat Tracking Error Detection}
\vspace{-0.25em}

We deliberately introduce a certain proportion of errors to the ground-truth beat annotations and ask the LLM to output the \textit{beat index range} containing beat errors based on the music performance data in the symbolic music format. We introduce three types of error on beat annotations: 1) insert an extra beat between adjacent beats; 2) delete a beat; and 3) offset the timing of one beat, where the offset should be greater than a 70ms tolerance \cite{davies2009evaluation}. In beat tracking tasks, error detection is not a binary classification problem per detected beat, because there are false negative predictions (i.e., missed beats error). Therefore, it is crucial to return the beat index range so that both false positive beats and missed beats can be captured.

As shown in Figure~\ref{fig:prompt-structure} (left), the music segment and the beat annotations with errors are provided in the JSON format. The notes and beats are sorted by the temporal positions (i.e., onsets or beat locations).

\subsubsection{Chord Extraction Error Detection}
\vspace{-0.25em}

We deliberately introduce chord annotation errors and ask the LLM to output the \textit{indices of incorrect chords} based on the music performance data in the symbolic music format. The chord errors are applied to the root, chord quality, and chord inversion attributes independently at a controlled rate.

As shown in Figure~\ref{fig:prompt-structure} (middle), we use the JSON format to represent the music segment and the chord annotations. The notes and chords are sorted by their temporal positions, and chords are notated as chord symbols in the conventional format \cite{harte2005symbolic}. 

\subsubsection{Key Estimation Error Detection}
\vspace{-0.25em}

We deliberately introduce key annotation errors and ask the LLM to output ``correct'' or ``incorrect'' based on the music performance data in symbolic format. The errors are introduced by selecting an incorrect key out of the other 23 major and minor keys randomly at a controlled rate.

As shown in Figure~\ref{fig:prompt-structure} (right), we use the JSON format to represent the music segment and key annotations, where the key annotation is given at the beginning. The predicted key is represented by a formatted string of tonic and mode (e.g., ``\texttt{A:min}'').

\subsection{Prompt Structure} \label{subsec:3.2}

Our investigation of prompt engineering methods indicates that a well-organized prompt structure is essential for successful MIR error detection. As shown in Figure~\ref{fig:prompt-structure}, the prompt of the three MIR tasks all consists of six components as follows:

\begin{itemize}[leftmargin=10pt]\setlength{\itemsep}{-2.5pt}

\vspace{-0.1cm}
\item \textbf{Background and Task Definition} introduces the MIR task and music domain background, and specifies the role of the LLM as a judge in assessing the correctness of MIR results.

\item \textbf{Music Concepts Description} introduces relevant music concepts about beat, chord, or key, together with examples of those concepts. For example, we show examples of chord root, quality, and inversion for chord extraction, to guide the LLM to better parse the chord labels such as \texttt{C:maj/3}.

\vspace{-0.1cm}
\item \textbf{Input Representation Explanation} specifies the data structure and format of the input music data. 

\vspace{-0.1cm}
\item \textbf{Techniques and Steps} provides reference steps to encourage the LLM to apply ``chain-of-thought'' in the error detection process. For example, we provide clear steps in the chord extraction task: 1) Identify Pitch Classes; 2) Determine Chord Quality; 3) Find Bass Note; 4) Consider Music Context; and etc.

\vspace{-0.1cm}
\item \textbf{Output Format Specification} defines the JSON-like output format, ensuring consistency for post-processing.

\vspace{-0.1cm}
\item \textbf{Data Input} provides the subject music piece and the MIR results to be judged, following the format defined in input representation explanation section.

\end{itemize}

\subsection{Concept Augmentation}\label{subsec:3.3}

The prompts defined in Section~\ref{subsec:3.2} contain extensive music concepts for each of the three tasks, which we regard as \textit{Basic Concepts}. Based on these, we apply \textit{concept augmentation} by either introducing new concepts or masking basic concepts in order to compare the LLM performance under varying amounts of music knowledge provided.

In \textit{Concept Introduction}, we add new concepts that are supposedly helpful for doing MIR tasks. For example, for beat tracking, we introduce ``rhythm'' to the LLM: we provide a brief description of on-beat notes and off-beat notes, and how to compute their density percentages. We explain how such concepts contribute to better judgments. 

Conversely, we also define the \textit{Concept Masking} operation, which eliminates or blurs music concepts at different levels. Such operations are used to examine the \textit{innate} reasoning ability of LLMs as a reference:

\vspace{-0.5em}
\begin{itemize}[leftmargin=10pt]\setlength{\itemsep}{0pt}

    \item \textbf{Music Attribute Masking}: removes explanations about music concepts pertaining to the musical objects under operations. For example, in chord extraction, ``root note'', ``chord quality'', and ``inversion'' are replaced by an abstract generic expression, ``a chord feature''. 

    \item \textbf{Task Masking}: on top of Music Attribute Masking, removes explanations about all concepts related to the MIR task, so that the LLM is required to reason about the correctness for a novel abstract task. For example, for beat tracking, the task becomes ``Please read a sequence of MIDI notes and music labels to determine the correctness of each label.'' All expressions that imply beat tracking will be deleted, including ``tempo'', ``fast'', ``slow'', etc., to ensure that the task information is not implied in any form.

    \item \textbf{Domain Masking}: on top of Task Masking, eliminates explanations about all concepts related to the \textit{music} domain to the greatest extent, leaving the LLM with an abstract logic-domain reasoning problem. For example, the LLM is told: ``You will be given some labels and the corresponding raw data. Your task is to tell me where the wrong labels are located?''

\end{itemize}

\section{Experiments}\label{sec:page_size}

\begin{figure*}[htbp]
    \centering
    \includegraphics[width=1\linewidth]{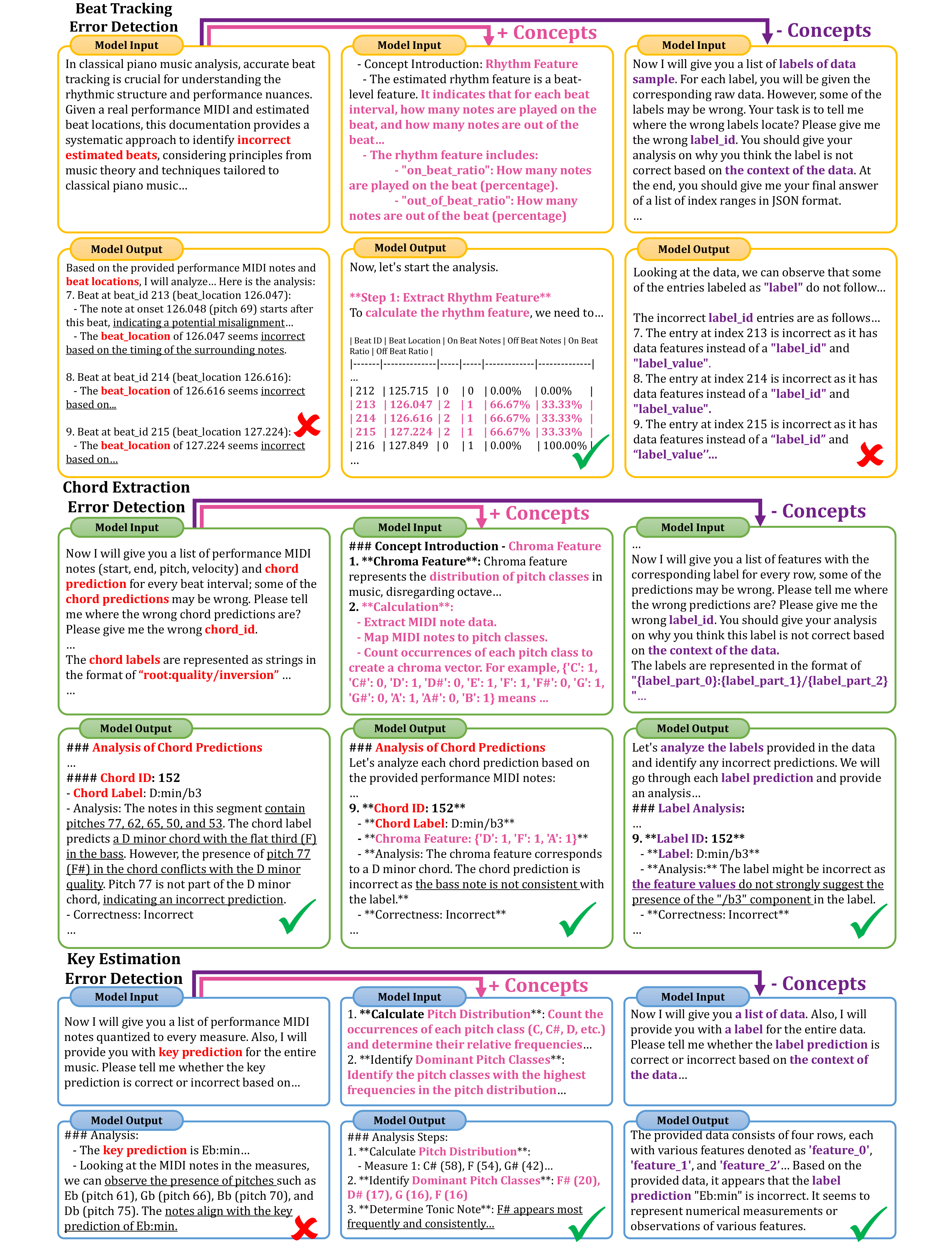}
    \caption{The impact of concept augmentation on GPT's behavior in three MIR error detection tasks: 1) \textit{Basic Concepts} (left), 2) \textit{Concept Introduction} (middle), and 3) \textit{Concept Masking}: all music domain concepts removed (right). Red color indicates the basic concepts. Pink color indicates the introduced concepts. Purple color represents the expression after masking all music-related concepts. Underlines denote reasoning process. The checkmark indicates a correct judgment made by GPT, while the cross indicates an incorrect judgment by GPT.}
    \label{fig:enter-label}
\end{figure*}

We conduct our LLM-based MIR tasks with GPT-3.5. We introduce the datasets in Section~\ref{subsec:4.1} and the evaluation metrics in Section~\ref{sec-metrics}. The evaluation results are provided in Section~\ref{subsec:4.3}.

\subsection{Datasets}
\label{subsec:4.1}
We use symbolic performance MIDI dataset for the three proposed tasks. For beat tracking error detection, we use classical piano recordings from the MAPS database \cite{maps}, specifically from the ``ENSTDkCl'' subset (29 pieces) which has been commonly used as a beat tracking test set. We also use the corresponding metrical annotations from \cite{amaps}. We randomly create beat errors by inserting (9\%), deleting (12\%), or offsetting (9\%) beats, resulting in an emulated MIR prediction with an F-score of 0.8370.

For both chord extraction and key estimation error detection, we use MIDI for Chinese pop songs on a subset of the POP909 dataset \cite{wang2020pop909}. 
For chord extraction, we randomly introduce errors in root, quality, or inversion with a ratio of 30\%, resulting in an ``MIR'' accuracy of 0.7327. We choose 50 songs and divide each song into segments with 32 chord labels. For key estimation, we test on 757 songs in the dataset whose ground-truth key is unchanged throughout the piece. We randomly select three four-measure segments for each song and modify 30\% of the key labels at random. A summary of the data statistics is shown in Table~\ref{tab:data_stats}.

\vspace{-0.5em}
\begin{table}[h]
\centering
\footnotesize
\setlength{\tabcolsep}{4pt}
\begin{tabular}{@{}cccc@{}}
\hline
 & Beat Tracking & Chord Extraction & Key Estimation \\ \hline
 \#Notes & 70,607 & 48,919 & 177,535 \\
 \#Labels & 14,194 & 9,200 & 2,271 \\
 \#Tokens (per call) & 6065.31 & 9256.80 & 3214.63 \\ 
 \hline
\end{tabular}
\vspace{-0.9em}
\caption{Statistics of the music data used for evaluation. The row \#Notes represents the total number of MIDI notes processed for each task. The row \#Labels indicates the number of labels used in the evaluation of each task. The row \#Tokens (per call) shows the average number of tokens per call fed into the GPT-3.5 model for each task.}
\label{tab:data_stats}
\end{table}

\begin{table*}[t!]
    
    \centering
    \footnotesize
    \renewcommand{\arraystretch}{0.98}
    \begin{subtable}[t]{1.0\textwidth}
        \centering
        \begin{tabular}{p{0.37\linewidth} p{0.11\linewidth} p{0.11\linewidth} p{0.11\linewidth} p{0.11\linewidth}}
            \hline
              Concept Augmentation & CPR$\uparrow$ & EDR$_{P}$$\uparrow$ & EDR$_{N}$$\uparrow$ & WS$\uparrow$ \\
            \hline
            Basic Concepts & 0.6681 & 0.3728 & 0.1794 & 0.5607 \\
            \hspace{0.2cm} + ``Rhythm'' & \textbf{0.8533} & 0.1496 & 0.0968 & \textbf{0.6520} \\
            \hspace{0.2cm} - ``Beat Location''(Music Attribute Masking) & 0.6898 & 0.3720 & 0.2008 & 0.5792 \\
            \hspace{0.2cm} - ``Beat Tracking''(Task Masking) & 0.5998 & 0.4010 & 0.2862 & 0.5296 \\
            \hspace{0.2cm} - ``Music''(Domain Masking) & 0.2418 & \textbf{0.7657} & \textbf{0.7061} & 0.3785 \\
            \hline
            Random & 0.513 $\pm$ 0.0586 & 0.4891 $\pm$ 0.0564 & 0.3238 $\pm$ 0.0608 & 0.4843 $\pm$ 0.0274 \\
            \hline
        \end{tabular}
        \vspace{-0.72em}
        \caption{Evaluation results on beat tracking error detection}
        \label{subtab:metrics-beat-tracking}
    \end{subtable}
    \hfill
    \begin{subtable}[t]{1.0\textwidth}
        \centering
        \begin{tabular}{p{0.44\linewidth} p{0.13\linewidth} p{0.13\linewidth} p{0.13\linewidth}}
            \hline
            Concept Augmentation & \textit{p} $\uparrow$ & \textit{r} $\uparrow$ & \textit{f} $\uparrow$ \\
            \hline
            Basic Concepts & 0.6345  &  0.6948  &  0.6207 \\
            \hspace{0.2cm} + ``Chroma'' & \textbf{0.6996}  &  \textbf{0.7174}  &  0.6290 \\
            \hspace{0.2cm} - ``Root''; ``Quality''; ``Inversion''(Music Attribute Masking) & 0.6503  &  0.6992  &  0.6376 \\
            \hspace{0.2cm} - ``Chord Extraction''(Task Masking) & 0.6497  &  0.6947  &  0.6362 \\
            \hspace{0.2cm} - ``Music''(Domain Masking) & 0.6848  &  0.7144  &  \textbf{0.6480} \\
            \hline
            Random & 0.5812 $\pm$ 0.0032 & 0.5003 $\pm$ 0.0034 & 0.5213 $\pm$ 0.0033 \\
            \hline
        \end{tabular}
        \vspace{-0.72em}
        \caption{Evaluation results on chord extraction error detection}
        \label{subtab:metrics-chord-extraction}
    \end{subtable}
    \hfill
    \begin{subtable}[t]{1.0\textwidth}
        \centering
        \begin{tabular}{p{0.44\linewidth} p{0.13\linewidth} p{0.13\linewidth} p{0.13\linewidth}}
            \hline
            Concept Augmentation & \textit{p} $\uparrow$ & \textit{r} $\uparrow$ & \textit{f} $\uparrow$ \\
            \hline
            Basic Concepts & 0.5789 & \textbf{0.6513} & 0.5965 \\
            \hspace{0.2cm} + ``Scale'' & 0.5847 & 0.6169 & \textbf{0.5972} \\
            \hspace{0.2cm} - ``Tonic''; ``Mode''(Music Attribute Masking) & 0.5754 & 0.5812 & 0.5782 \\
            \hspace{0.2cm} - ``Key Estimation''(Task Masking) & 0.5840 & 0.6143 & 0.5960 \\
            \hspace{0.2cm} - ``Music''(Domain Masking) & \textbf{0.5927} & 0.4161 & 0.4085 \\
            \hline
            Random & 0.5779 $\pm$ 0.0086  &  0.4977 $\pm$ 0.0093  &  0.5186 $\pm$ 0.0089 \\
            \hline
        \end{tabular}
        \vspace{-0.72em}
        \caption{Evaluation results on key estimation error detection}
        \label{subtab:metrics-key-estimation}
    \end{subtable}
    \vspace{-1.4em}
    \caption{The evaluation results of GPT on three MIR error detection tasks: beat tracking, chord extraction, and key estimation. Each task is assessed under different concept augmentation. ``+'' denotes \textit{Concept Introduction}. ``--'' denotes \textit{Concept Masking}. $\uparrow$ indicates that higher values are better. $p$, $r$, and $f$ stand for precision, recall, and F-score, respectively. 
    }
    \label{tab:metrics-all-tasks}
\end{table*}

\vspace{-1.2em}
\subsection{Evaluation Metrics}\label{sec-metrics}

We design metrics to evaluate the performance of LLMs in identifying errors in MIR annotations. Since our approach does not directly predict MIR annotations, our metrics differ from existing MIR evaluation metrics. For chord extraction and key estimation tasks, we regard error detection as a binary classification task in which each chord or key label is classified as correct or incorrect. We use weighted precision, recall, and F1-score to evaluate GPT's performance on both correct and incorrect classes \cite{sklearn}.

In beat tracking error detection, the beat sequence with potential errors is typically not one-to-one aligned with the ground truth beats. We consider three types of beat locations: 1) correctly identified beats, 2) additional beats, and 3) missing beats, which are also referred to as true positives, false positives, and false negatives, respectively, in conventional beat tracking tasks \cite{davies2009evaluation}. We use TP, FP, and FN to denote these sets of beat positions and $I$ to denote the union of time intervals predicted by an LLM error detector. We consider the following metrics:  

\begin{itemize}[leftmargin=10pt]\setlength{\itemsep}{0pt}
    \item \textbf{CPR (Correct Pass Rate on TP)} is defined as $\frac{|\text{TP} - I|}{|\text{TP}|}$, which
    measures the proportion of true positives that are correctly identified (by GPT) as ``correct beats''.
    
    \item \textbf{EDR$_{P}$ (Error Detection Rate on FP)} is defined as $\frac{|I \cap \text{FP}|}{|\text{FP}|}$
    , which evaluates the proportion of false positives that are correctly identified (by GPT) as ``incorrect beats''.
    
    \item \textbf{EDR$_{N}$ (Error Detection Rate on FN)} is defined as: $\frac{|I \cap \text{FN}|}{|\text{FN}|}$, which evaluates the proportion of false negatives that are correctly identified (by GPT) as ``incorrect beats''.
\end{itemize}
Finally, we compute a weighted average of these metrics, denoted by WS:

\begin{equation}
\text{WS} = \frac{\text{CPR} \times \text{${|\text{TP}|}$} + \text{EDR}_{\text{P}} \times \text{${|\text{FP}|}$} + \text{EDR}_{\text{N}} \times \text{${|\text{FN}|}$}}{\text{${|\text{TP}|}$} + \text{${|\text{FP}|}$} + \text{${|\text{FN}|}$}}\text{.}
\end{equation}

\subsection{Evaluation Results}\label{subsec:4.3}

We evaluate the performance of GPT on three MIR error detection tasks. We first use the prompt with Basic Concepts and compare it with a random baseline, as well as prompts under different concept augmentation methods (see Section~\ref{subsec:3.3}). The results are summarized in Table~\ref{tab:metrics-all-tasks}.

The results of beat tracking error detection task are shown in Table~\ref{subtab:metrics-beat-tracking}. The random baseline is implemented by first randomly selecting $k$ beat labels and joining consecutively selected beats into time intervals serving as detected error ranges. In Concept Introduction, we guide the GPT to compute the number of on-beat and off-beat note percentages, and in Concept Masking, we apply music attribute, task, and domain masking incrementally. Results show the basic prompt outperforms the random baseline in all prompt settings. Moreover, as the number of concepts decreases, the performance of GPT in judging the correctness of beat labels shows an overall downward trend. 

The results of chord extraction error detection task are shown in Table~\ref{subtab:metrics-chord-extraction}. The random baseline detects incorrectness with a probability of 50\%. In Concept Introduction, we show GPT the chord chroma concept and encourage GPT to deduce the pitch distribution from input music. Results show that all GPT settings far exceed the random baseline. There remains a downward trend as the number concepts decreases except in the Domain Masking setting.

The results of key estimation error detection task are shown in Table~\ref{subtab:metrics-key-estimation}. The random baseline and concept augmentation are implemented similarly to those of chord extraction. In Concept Introduction, we show GPT the scale concept. Results show that GPT performs slightly better than the random baseline in F-score and recall, and similar to the baseline in precision. The downward trend of concept augmentation is less salient.

Finally, we provide a case study (Figure~\ref{fig:enter-label}) to illustrate GPT's behavior under different settings of concept augmentation. In all tasks, GPT exhibits general time series analysis abilities even when music concepts are all masked, and the introduced music concepts help GPT to reason in a more musical fashion, particularly in beat tracking. However, we also observe limitations, including high randomness in output, sensitivity to prompts, and hallucination \cite{hallucination}. These issues make it challenging to empirically summarize or conjecture GPT's reasoning abilities in solving MIR problems in general.

\section{Conclusion and Future Work}
\label{sec:typeset_text}

In conclusion, we have proposed a methodology to solve MIR problems with text-based LLMs with prompt engineering. We evaluate the performance of GPT-3.5 in error detection across three MIR tasks and find out that GPT's music reasoning ability in MIR tasks can be enhanced when provided with well-structured prompts with music concepts. Across all three MIR error detection tasks, GPT consistently outperforms random baseline methods and demonstrates improved performance when prompted with additional music knowledge. In this study, we establish a baseline for assessing LLMs' ability to understand music solely through reasoning, paving the way for future LLM-based MIR research. In the future, we will consider evaluating LLMs' judging ability on real MIR errors instead of synthetic ones and using fine-tuning techniques to better explore LLM-based MIR study.

\section{Acknowledgments}

This research has been supported by the Social Sciences and Humanities Research Council of Canada (SSHRC 895-2022-1004) and the China Scholarship Council.

\bibliography{ISMIR2024_template}

\end{document}